\documentstyle[12pt,aps,epsf]{revtex}
\newcommand{\ep}{\varepsilon}
\begin{document}
\rightline{Preprint YerPhI 1512(12)-98}
\begin{center}
{\Large {\bf Optical Constants and Functions of Corundum Single Crystals
in the  Vacuum Ultraviolat Region}}\\[1cm]
{\large V.V. Harutunyan, T.S. Hakobyan
A.S.Hovhannesian,\\
V.A. Gevorkyan,
N.E. Grigoryan, A.K.Avakyan}\\[5mm]
{\it Yerevan Physics Institute, 2 Alikhanian Br.\\
       375036 Yerevan Armenia}\\[1cm]
{\large V.A.Grigoryan}\\[5mm]
{\it Yerevan State University, 1 Manoogian\\
375049 Yerevan Armenia}
\end{center}
\vspace{1cm}
\begin{abstract}

Optical reflection spectra of corundum single crystals grown by HOC
and Verneil  methods are investigated before and after irradiated
by electrons and neutrons using synchrotron radiation polarization
(5-30eV).

The peculiarities of defect formation and compositions of surface
electron structure and the main tendencies  of atomic structure
change due to surface destruction in irradiated corundum crystals
are considered.

Using the Kramers-Kronig analysis the spectra of optical constants and
functions are obtained for nonirradiated-irradiated corundum crystals.
The plasmon energy and the energy losses
also depend on the orientation of $\vec{E}$ to $C_3$. The effects observed
are due to changes in the density of states of the $2p$-band electrons.
\end{abstract}

\section{Introduction}

Aluminum oxide $\alpha-Al_2O_3$ is of great importance to modern technology.
It is using as crystal optical materials, active laser element matrices,
radiation stable ceramics, e.t.c.

Recently there has been a considerable interest in studding
of the optical properties of corundum and the radiation influence
on its electron-energetic structure in ultra violate (UV) and vacuum
ultra-violate (VUV) spectra regions [1-8].
Corundum in $\alpha$-phase is
transparent in VUV spectrum region. It conserves this property under
high energy particle radiation.

The phenomenological picture of wave propagation in matter is often
used to study the interaction of electromagnetic emission with condensed
matter states \cite{9}. The ideal conditions for all phenomenological
constants are imposed (proposed). Among them are dielectric
$\varepsilon$ and optical $n$, $\kappa$ constants. The studding of
dielectric and other
constants gives the direct information  about intrinsic processes in
crystals, in particular, corundum. However by some reasons the direct
measurement of these constants is difficult task. For example,
the very thin matter layers are necessary in order to measure the
$\alpha$-absorption ($\approx 10^{-5}cm^{-1}$). So, the indirect
measurement of optical constants is one of the main problems
of the spectroscopy.

There is a sufficiently accurate method  to determine the optical
constants by means of reflection coefficient of light in wide band
region of spectra.
One of the main methods applying in the fundamental absorption
region is
the dielectric constant calculation using Kramers-Kronig
relations \cite{10}.

In this article the  optical constants and optical functions
of corundum monocrystal are  investigated  using the optical
reflection spectra and irradiation influence on their
energetic structure.

\section{The methods of investigations}

The relative reflection of crystalls was measured by means of synchrotron
radiation (SR) of C-60 accelerator. The device used in experiment is based on
monochromator constructed by Vodsvort modification scheme without band,
which has the better resolution than 10nm. The resolution of
spectra measurement in low energy region  was
$\pm 5\%$, in high energy region it was
$\pm3\%$ and energy resolution was about $\pm0.1eV$.

The dispersion relations, $\varphi$ phase, $n$ and $k$ constants was
calculated using the experimental data of reflection coefficient $R(E)$ at
$300K$ and Kramers-Kronig dispersion relation.
The reflection coefficient is known in the restricted frequency region
from $\omega_{min}$ to $\omega_{max}$.  Thus the contribution
of other regions is essential.
The frequencies below $\omega_{min}$ are in the
transparency region of $\alpha-Al_2O_3$ crystal.
At the same  time $\varphi=0$, because $r$ is real and the reper values
of $R(E)$ was determined from $n$ coefficients, which was measured
independently using Frenel formulae. The high frequency approximation
($E>30eV$) is $R(E)\sim\omega^{-4}$.

The optical functions $n$, $k$, $\ep_1$, $\ep_2$, $Im(\ep)^1$,
$Im(\ep+1)^1$, $N_{eff}$ [10-12]
were determined from $\Theta(E)$
in $5-30eV$ region, where $R(E)$ was measured.

For $\alpha-Al_2O_3$ the molecule density is $2.3\times10^{22}cm^{-3}$.

For the $R(E)$ measurement the $\alpha-Al_2O_3$ samples, which were
grew by means of horizontally oriented crystallization (HOC) and
Verneil methods,
was used. The optical axes $C_3$ was parallel to  large side of crystal.
The samples had a good mirror surface.
The crystals were cultivated by $HCl$ and washed by pure water before the
reflection spectra measurements.

The corundum monocrystals were irradiated by electrons of $50MeV$ energy
at $273K$ and by reactor neutrons of energy $2MeV$ at $373K$.

\section{Result discussion.}

Fig.1 and Fig.2 contain the experimental reflection spectra $R(E)$
of HOC and Verneil crystals. It can be observed from the pictures that
a weak maximums (bands) of $5.4eV$; $6.3eV$; $7eV$ for Verneil and
of $6.1eV$ for HOC crystals appear in low energy band of spectra $R(E)$.

These bands for irradiated crystals we related to $F^{\pm},F^{2+}$
and $F$ centers \cite{13}. The appearance of bands $6-6.5eV$ and $7-7.4eV$
is caused  by ferrum group ions $Cr,Mn,Fe,Ti,Ni,Co$, which have different
anisotropy.
 $Cr^{3+}$ ion concentration in Verneil samples was $\sim10^{17}cm^{-3}$,
in HOC samples was $\sim10^{15}cm^{-3}$.

After radiation the $R(E)$ spectra maximums, mentioned above, increase and
become more precise (the curves).
These are the absorption bands of $5.4eV$; $6.3eV$; $7eV$ and $8eV$,
caused by $F^+$-center and the band of $6.1eV$, caused by $F$-center.

The radiation and essentially electron, neutron action
changes the reducible center medium stronger than it is
expected by the elastic interaction theory.

Note, that the  defect formation   under the  neutron radiation
of corundum is much more faster than one under the high speed
electron radiation at same doses.
It can be seen from the experimental data of $R(E)$ that the
electric field changes intensively on near-surface layers due to
defect compound. This affects on point color centers.

In $R(E)$ spectra of $\alpha-Al_2O_3$ the intensive long wave peak
at $8.9-9.1eV$ appears. The exciton peak depends on SR polarization
and orientation of crystal (Fig.1,2). After the radiation the shift of
exciton band maximum on $\pm0.2eV$ can be observed. The anion
$\Gamma$-excitons
are created, because the high valent band of $Al_2O_3$ is formed by
$2p$ states of oxide anions \cite{14}. It is known that the ...
of exciton bands is the consequence of the selection rule, according
to which the during the absorption photon quasimomentum is conserved
and translated to exciton.

The large bandwidth and inhomogeneosy of exciton spectra absorption
is explained by mixed (hybrid) character of exciton states.
It is known that that the hybrid excitons as a "sum" of hyperbolic and
parabolic ones are observed at the energetical degeneracy of critical
points $M_0$ and $M_1$.
Hyperbolic excitons are not dynamically stable and decompose on free
electrons and holes. The mixed character of exciton absorption bands
can be explained by different bond energies of excitons $M_0$ and $M_1$.

There is a nonelementar wide-band maximum in the reflection spectra of
irradiated crystals in the region $12-25eV$. Band group in $R(E)$
of HOC crystals is more precise than in Verneil samples. This fact
is connected with the presence of doped ions on the surface
of Verneil crystals, which leads to the stoichiometry violation in anion and
cation sublattices.

From the pictures, the values of reflection coefficients in  radiated
samples begin to fall. It proves the radiation influence on crystal
surface.  Besides the near-surface defects the scattering effect
increases too, which also contributes to reflection coefficient.

During the (electron, neutron) radiation of corundum the
crystal structure disorder of near-surface increases because of
high concentration of vacancies  and interband ions. The radiation
stimulated desorbation process takes place with recovered $Al$ phase as
a result \cite{5,6}.

Besides the point vacancy defects the complicated color centers
of $[Al_iF]$ type with various charge states can be formed \cite{14}.

Other mechanism of defect formation in near-surface layers is also
possible. It is based on dislocation formation model \cite{15,16}.

The consequence of corundum surface decoration by $Al$ ions
is the reflection improvement of radiated crystals at energy
$\hbar\nu>25eV$.

The optical constants, calculated from experimental $R(E)$ changes
also (Fig.3-8). The band group is well observed in $n,\kappa,\ep_1,\ep_2$
spectra also.

It is easy to see from Fig.3,4 that the optical constant $n$
has thin structure at different orientations between $\vec{E}$ and $C_3$.
The high resolution reflection spectra
was measured owing to both higher intensity and polarization of SR than
in \cite{10,11}.
The difference in $n,\kappa$ spectra on Fig.3 and 4 is conditioned by
the value of special energetic state of surface, which affects on
radiation irregularity formation  in HOC and Verneil samples.

The reflection band group (Fig.1,2)in region $12-25eV$ is showed
precisely in $n,\kappa,\ep_1,\ep_2$ spectra also (Fig.3-8).
According to calculations, the values of $\ep_1$ and $\ep_2$ differs
for different orientations between $\vec{E}$ and $C_3$. The radiation
influence on dielectric constant is essential. Up to energy $8.2eV$
$\ep_2$ and $\kappa$ vanish. It is easy to see from the $\ep_2$ spectra,
that for HOC in case of $E\bot C_3$ the enegry is $23eV$ and in case
of $E||C_3$ it is $21eV$.
For Verneil samples the $\ep_2$ value corresponds to $23eV$ and $20eV$
and after the electron radiation it becomes $21eV$ and $21.5eV$.

The data, reduced from optical constant spectra (Fig.5,8), confirm
that the polarization effects are connected with splitting of valence
$2p$-states of oxygen in low symmetry fields of corundum lattice.

The interband state maximal density change effect probably is connected
with the transitions from the highest intensive maximum  in density of
states of $2p$-band. This is due to increase of defect concentration
in near-surface layers (expessially for neutron radiation).

In order to estimate quantitatively the interband intervals the
calculations of interband combined density $\ep_2E^2$ (Fig.9,10)
and energetic loss functions $Im(\ep)$ and $Im(\ep+1)^{-1}$
(Fig.11,12).

It can be seen from the  energetic loss functions of electrons in
HOC crystals that in case of $E\bot C_3$
the largest value of volumetric plasmon
$Im(\ep)^{-1}$ of radiated and irradiated crystal are $25eV$ and
$24.5eV$ correspondingly (the curves 1,3).
The maximums of surface plasmon
$Im(\ep+1)^{-1}$ of radiated and irradiated crystal are $24eV$ and
$23.5eV$ correspondingly (the curves 2,4).
In case of $E||C_3$ for $Im(\ep)^{-1}$ one has maximums at $25eV$, $23.5eV$
(the curves 1,3) and for $Im(\ep+1)^{-1}$ one has maximums at
$23.5eV$, $23eV$.

For Verneil crystals in the energetic loss functions the maximums
in case of $E\bot C_3$ (Fig.12)
for $Im(\ep)^{-1}$ are $26eV$ and $24.5eV$ (the curves 1,3) and for
$Im(\ep+1)^{-1}$ its are $25eV$ and $24.5eV$ (the curves 2,4).
In case of $E\parallel C_3$ (Fig.12) for $Im(\ep)^{-1}$ one has
maximums at $25.5eV$, $24.5eV$ (the curves 1,3) and for
$Im(\ep+1)^{-1}$ one has maximums at $25eV$, $24eV$.

Note, that the value $Im(\ep)^{-1}$ is proportional to the energy loss
probability of quasi-free gas electrons in $\alpha-Al_2O_3$. It describes
the maximal fluctuation of plasmon energy, which constitutes  $26eV$ at
nonpolarized light.

Besides the intensive plasma fluctuations in $Im(\ep)^{-1}$ spectra
the weak peeks are observed in the regions $9.8$,  $13$, $14.5$, $15.5$,
$20$, $21.5eV$. Its are the consequences of energy loss of near-surface
and volumetric $Al$ plasmons of nonirradiated and irradiated corundum crystals.

The characteristic loss spectra shows that $Al$ plasmon peek location
depends drastically on both type and doze  of radiation action and
leads to shift on $0.2-0.5eV$.

The effective electron number $N_{eff}$ is calculated from the sum rule,
which express the fact that the sum of all oscillator forces
equals to electron number.

In the plasmon energy region $N_{eff}$ reaches $14$ electrons per $Al_2O_3$
molecule, which has $18$ valence $2p$-electrons.
The fact that its are not saturated means the oscillator
are not entirely strengthless. Hence, the absorption, which contributes to
the sum rule,  must be below and higher $30eV$ (Fig.13).

Using Arand curves for HOC and Verneil crystals the resonance energies of
partial oscillators have been precisely determined. Every semicircle
on Fig.14-17 corresponds to one oscillator, i.e. one interband transition.

To calculate the resonance energies one first have extrapolated every
semicircle and then have found the values $O_i$ using maximum
$\ep_\alpha(E)$.

In tables 1,2 the identifications of interband transitions for HOC
and Verneil crystals are presented for two polarizations
($\vec{E}\bot C_3,\vec{E}||C_3$) taking into account
the electron energetic structure of corundum.

Thus, our identification of interband transition for corundum crystals shows
two effects of polarization and radiation. The first one is connected
with splitting of the 2d valence electrons of oxygen in the low symmetry
crystal field of corundum. The second one is related to transition in
the upper intensive  maximum of the 2p band state density after
increasing the
radiation defects in near-surface layers.

\begin{table}[ht]
\caption{ HOC nonirradiated and irradiated by neutrons }
\vspace{5mm}
 \begin{tabular}{|ll|ll|}
\hline
 $O_1^\bot $ -- $O_1^\parallel$ & 9.20 -- 9.0 eV &
 $O_{1,rad}^\bot $ -- $O_{1,rad}^\parallel$ & 8.60 -- 9.0 eV \\
\hline
 $O_2^\bot$ -- $O_2^\parallel$ & 11.60 -- 10.0 eV &
 $O_{2,rad}^\bot$ -- $O_{2,rad}^\parallel$ & 9.0 -- 9.4 eV \\
\hline
 $O_3^\bot$ -- $O_3^\parallel$ & 12.60 -- 12.80 eV &
 $O_{3,rad}^\bot$ -- $O_{3,rad}^\parallel$ & 12.80 -- 12.60 eV \\
\hline
 $O_4^\bot$ -- $O_4^\parallel$ & 18.40 -- 18.60 eV &
 $O_{4,rad}^\bot$ -- $O_{4,rad}^\parallel$ & 18.40 -- 19.20 eV \\
\hline
 $O_5^\bot$ -- $O_5^\parallel$ & 21.80 -- 21.0 eV &
 $O_{5,rad}^\bot$ -- $O_{5,rad}^\parallel$ & 21.40 -- 21.20 eV \\
\hline
 $O_6^\bot$ -- $O_6^\parallel$ & 22.0 -- 21.40 eV &
 $O_{6,rad}^\bot$ -- $O_{6,rad}^\parallel$ & 22.80 -- 22.20 eV \\
\hline
 $O_7^\bot$ -- $O_7^\parallel$ & 23.0 -- 23.20 eV &
 $O_{7,rad}^\bot$ -- $O_{7,rad}^\parallel$ & 26.60 -- 24.0 eV \\
\hline
 $O_8^\bot$ -- $O_8^\parallel$ & 24.60 -- 24.40 eV &
 $O_{8,rad}^\bot$ -- $O_{8,rad}^\parallel$ & 28.40 -- 25.60 eV\\
\hline
\end{tabular}
\end{table}

\noindent
1-3 [$O^\bot$ -- $O^\parallel$ ] -- [8.60--13.0] eV -- 2p[$O^2$];\\
4-5 [$O^\bot$ -- $O^\parallel$ ] -- [18.0--20.5] eV -- 3s[$Al^{3+}$]
-2s[$O^{2-}$];\\
5-8 [$O^\bot$ -- $O^\parallel$ ] -- [21.0--28.5] eV -- 3sp$^2$[$Al^{3+}$]
-2s[$O^{2-}$].

\begin{table}[ht]
\caption{ Verneyl: nonirradiated and irradiated by electrons }
\vspace{5mm}
 \begin{tabular}{|ll|ll|}
\hline
 $O_1^\bot $ -- $O_1^\parallel$ & 9.40 -- 9.20 eV &
 $O_{1,rad}^\bot $ -- $O_{1,rad}^\parallel$ & 9.60 -- 9.20 eV \\
\hline
 $O_2^\bot$ -- $O_2^\parallel$ & 11.0 -- 11.30 eV &
 $O_{2,rad}^\bot$ -- $O_{2,rad}^\parallel$ & 11.20 -- 11.60 eV \\
\hline
 $O_3^\bot$ -- $O_3^\parallel$ & 12.20 -- 12.0 eV &
 $O_{3,rad}^\bot$ -- $O_{3,rad}^\parallel$ & 12.0 -- 11.80 eV \\
\hline
 $O_4^\bot$ -- $O_4^\parallel$ & 16.80 -- 17.0 eV &
 $O_{4,rad}^\bot$ -- $O_{4,rad}^\parallel$ & 19.40 -- 19.20 eV \\
\hline
 $O_5^\bot$ -- $O_5^\parallel$ & 19.80 -- 18.80 eV &
 $O_{5,rad}^\bot$ -- $O_{5,rad}^\parallel$ & 20.20 -- 20.0 eV \\
\hline
 $O_6^\bot$ -- $O_6^\parallel$ & 21.0 -- 19.20 eV &
 $O_{6,rad}^\bot$ -- $O_{6,rad}^\parallel$ & 21.0 -- 21.20 eV \\
\hline
 $O_7^\bot$ -- $O_7^\parallel$ & 21.40 -- 20.40 eV &
 $O_{7,rad}^\bot$ -- $O_{7,rad}^\parallel$ & 22.80 -- 21.60 eV \\
\hline
 $O_8^\bot$ -- $O_8^\parallel$ & 22.60 -- 22.20 eV &
 $O_{8,rad}^\bot$ -- $O_{8,rad}^\parallel$ & 24.80 -- 23.80 eV\\
\hline
 $O_9^\bot$ -- $O_9^\parallel$ & 23.40 -- 24.0 eV &
 $O_{9,rad}^\bot$ -- $O_{9,rad}^\parallel$ &  -- 24.80 eV \\
\hline
 $O_{10}^\bot$ -- $O_{10}^\parallel$ & 26.0 -- 25.80 eV &
 $O_{10,rad}^\bot$ -- $O_{10,rad}^\parallel$ & --- \\
\hline

\end{tabular}
\end{table}

\noindent
1-3 [$O^\bot$ -- $O^\parallel$ ] -- [9.0--12.5] eV -- 2p[$O^2$];\\
4-5 [$O^\bot$ -- $O^\parallel$ ] -- [16.5--20.5] eV -- 3s[$Al^{3+}$]
-2s[$O^{2-}$];\\
6-10 [$O^\bot$ -- $O^\parallel$ ] -- [21.0--26.0] eV -- 3sp$^2$[$Al^{3+}$]
-2s[$O^{2-}$].

\newpage
\section{Figure captions}

\begin{figure}[ht]
%\epsfxsize=0.25\textwidth
%\epsfbox{p1.ps}
%\epsfbox{p2.ps}
	\caption{\label{fig:1} Reflection spectra vs. Polarization of SR with respect to the optical axis of
 corundum crystals (HOC) $C_3$. (1)-nonirradiaed; (2)-irradiated dose $10^{17}n/cm^2$}
\end{figure}

\begin{figure}[ht]
%\epsfxsize=0.25\textwidth
%\epsfboxp{3.ps}
%\epsfboxp4.ps}
        \caption{\label{fig:2} Reflection spectra vs. Polarization of SR with respect to the optical axis of
 corundum crystals (Verneil) $C_3$. (1)-nonirradiaed; (2)-irradiated dose $3\cdot 10^{17}el/cm^2$}
\end{figure}

\begin{figure}[ht]
%\epsfxsize=0.45\textwidth
%\epsfboxp5.ps}
%\epsfboxp6.ps}
\caption{\label{fig:3}  Optical constants of corundum single crystals, (HOC):
          1({\bf n}), 3({\bf k})-nonirradiated; 2({\bf n}), 4({\bf k})-irradiated dose $10^{17}n/cm^2 $.}
\end{figure}

\begin{figure}[ht]
%\epsfxsize=0.45\textwi0dth
%\epsfboxp7.ps}
%\epsfboxp{8.ps}
\caption{\label{fig:4} Optical constants of corundum single crystals, (Verneil):
          1({\bf n}), 3({\bf k})-nonirradiated; 2({\bf n}), 4({\bf k})-irradiated dose $3 \cdot 10^{17} el/cm^2 $.
 }
\end{figure}

\begin{figure}[ht]
%\epsxsize=0.45\textwidth
%\epsbox{p9.ps}
\caption{\label{fig:5} Dielectric constants of corundum single crystals (HOC):
          1($\epsilon_1$), 3($\epsilon_1$)((2)-nonirradiated; 2($\epsilon_1$)
(1), 4($\epsilon_1$) (2)-irradiated dose $10^{17}n/cm^2$.}
\end{figure}

\begin{figure}[ht]
%\epsfxsize=0.45\textwidth
%\epsfbox{p10.ps}
\caption{\label{fig:6} Dielectric constants of corundum single crystals (HOC):
          1($\epsilon_1$), 3($\epsilon_2$)-nonirradiated; 2($\epsilon_1$), 4($\epsilon_2$)
irradiated dose  $10^{17}n/cm^2$.
}
\end{figure}

\begin{figure}[ht]
%\epsfxsize=0.45\textwidth
%\epsfbox{p11.ps}
\caption{\label{fig:7} Dielectric constants of corundum single crystals (Verneil):
          1($\epsilon_1$), 3($\epsilon_2$)-nonirradiated; 2($\epsilon_1$), 4($\epsilon_2$)-irradiated dose $3 \cdot 10^{17}el/cm^2$.
}
\end{figure}

\begin{figure}[ht]
%\epsfxsize=0.45\textwidth
%\epsfbox{p12.ps}
\caption{\label{fig:8} Dielectric constants of corundum single crystals (Verneil):
          1($\epsilon_1$), 3($\epsilon_2$)-nonirradiated; 2($\epsilon_1$), 4($\epsilon_2$)-irradiated dose $3 \cdot 10^{17}el/cm^2$.
}
\end{figure}

\begin{figure}[ht]
%\epsfxsize=0.45\textwidth
%\epsfbox{p13.ps}
\caption{\label{fig:9} Interband combined density ($\epsilon_2 E^2$)  (HOC): 1,3-nonirradiated;
          2,4-irradiated dose $ 10^{17}n/cm^2$.
}
\end{figure}

\begin{figure}[ht]
%\epsfxsize=0.45\textwidth
%\epsfbox{p14.ps}
\caption{\label{fig:10} Interband combined density ($\epsilon_2 E^2$)  (Verneil): 1,3-nonirradiated;
          2,4-irradiated dose $3 \cdot 10^{17}el/cm^2$.
}
\end{figure}

\begin{figure}[ht]
%\epsfxsize=0.45\textwidth
%\epsfbox{p15.ps}
%\epsfbox{p16.ps}
\caption{\label{fig:11} Energetic loss functions of corundum single crystals (HOC):
          1,2-nonirradiated; 3,4-irradiated dose  $10^{17}n/cm^2$.}
\end{figure}

\begin{figure}[ht]
%\epsfxsize=0.45\textwidth
%\epsfbox{p17.ps}
%\epsfbox{p18.ps}
\caption{\label{fig:12} Energetic loss functions of corundum single crystals (Verneil):
          1,2-nonirradiated; 3,4-irradiated dose  $10^{17}n/cm^2$.
 }
\end{figure}

\begin{figure}[ht]
%\epsfysize=0.45\textheight
%\epsfbox{p19.ps}
%\epsfysize=0.45\textheight
%\epsfbox{p20.ps}
\caption{\label{fig:13} Effective electron number of  $\alpha-Al_2O_3$ single crystals
        (HOC, Verneil):
                1,2-nonirradiated:  $\vec{ E} \parallel C_3$  and  $\vec{ E} \bot C_3$;
                3,4-irradiated:  $E \parallel C_3$  and  $\vec{E} \bot C_3$.}
\end{figure}

\begin{figure}[ht]
%\epsfxsize=0.45\textwidth
%\epsfbox{p21.ps}
%\epsfbox{p22.ps}
\caption{\label{fig:14} Argand diagram for corundum crystals (HOC, $\vec{E}\bot C_3$)
$a$-nonirradiated,
          $b$-irradiated  $10^{17}n/cm^2$.
}
\end{figure}

\begin{figure}[ht]
%\epsfxsize=0.45\textwidth
%\epsfbox{p23.ps}
%\epsfbox{p24.ps}
\caption{\label{fig:15} Argand diagram for corundum crystals (HOC, $\vec{E}\parallel
C_3$) $a$-nonirradiated, $b$-irradiated  $10^{17}n/cm^2$.
}
\end{figure}

\begin{figure}[ht]
%\epsfxsize=0.45\textwidth
%\epsfbox{p25.ps}
%\epsfbox{p26.ps}
\caption{\label{fig:16} Argand diagram for corundum crystals (Verneil, $\vec{E}\bot C_3$)
$a$-nonirradiated,
          $b$-irradiated $3 \cdot 10^{17}el/cm^2$.
 }
\end{figure}

\begin{figure}[ht]
%\epsfxsize=0.45\textwidth
%\epsfbox{p27.ps}
%\epsfbox{p28.ps}
\caption{\label{fig:17} Argand diagram for corundum crystals (Verneil, $\vec{E}\parallel
C_3$) $a$-nonirradiated,
          $b$-irradiated $3 \cdot 10^{17}el/cm^2$.
}
\end{figure}


\begin{thebibliography}{99}

\bibitem{1} La S.Y., Bartram R.H.,  Cox R.T., The $F^+$ center in
reactor-irradiated aluminium oxide, {\em J. Phys. Chem. Sol.} 34, p.1079,
1973

\bibitem{2} Lee, K.H., Crawford, J.H.,... Electron centers in single crystals
$Al_2O_3$, {\em Phys. Rev.} B15, p.4065, 1977

\bibitem{3} Kuznetsov, A.I., Ilmas, E.P., The reflection of $Al_2O_3$
crystals from  the sides (1011) and (0001) in the region $4.4-14.5eV$,
{\em FTT} 17, p.2132, 1975 (Russia)

\bibitem{4}
Tomiki, T., Futemma, T., Kato, H., at all, The optical spectra
of $\alpha-Al_2O_3$ single crystals in the vacuum ultraviolet region,
{\em J. Phys. Soc. J.} 58, p.1487, 1989

\bibitem{5}
Harutunyan, V.V., Gevorkyan, V.A., Eritsian, G.N.,
Parameters of interband transitions in corundum crystals,
{\em Phys. Stat. Sol.} 183, p.K23, 1994

\bibitem{6}
Harutunyan, V.V., Babayan, A.K., Gevorkyan, V.A., Makhov, V.N.,
The $\alpha$-irradiation influence on
defect formation on $\alpha-Al_2O_3$
monocrystal surface, Poverkhnost v.10-11, p.128, 1994 (Russia)


\bibitem{7}
Harutunyan, V.V., Belsky, A.N., Gevorkyan, V.A., Mikhailin, V.V.,
Eritsian, G.N.,
The short lifetime deffect formation in $\alpha-Al_2O_3$ under the
synchrotron radiation, {\em Radiation effects and deffects in solids}
134, p.243, 1995


\bibitem{8}
Harutunyan, V.V., Babayan, A.K., Gevorkyan, V.A., Sarkisyan, G.K.,
Makhov, V.N.,
The investigation of surface state of irradiated by Pb ions
corundum monocrystals by means of synchrotron radiation,
Poverkhnost, v.10, p.  ,1998  (Russia)


\bibitem{9}
Ternov, I.M., Mikhailin, V.V., The synchrotron radiation.
Theory and experiment, M., p.256, 1989 (Russia)

\bibitem{10}
Arakawa, E.T., Williams, M.W., Optical properties of aluminum
oxide in the vacuum ultraviolet, {\em J. Phys. Chem. Sol.} 29, p.735,
1968

\bibitem{11}
Ditchfield, R.W., Measurement and interpretation of the plasmon
energy in alumina, {\em Sol. St. Commun.} 19, pp.443, 1976

\bibitem{12}
Harutunyan, V.V., Gevorkyan, V.A., Grigoryan, N.E.,
Eritsian, G.N.,
The electron irradiation influence on $\alpha Al_2O_3$
monocrystal band structure, Preprint YerPhI-1183(60)-89,
Yerevan, 1989

\bibitem{13}
Harutunyan, V.V., Belsky, A.N., Gevorkyan, V.A., Grigoryan, N.E.,
Eritsian, G.N., Vacuum ultra-violat luminescence excitation
spectra of $Al_2O_3$ single crystals,
{\em Nucl. Instr. and Meth. Phys. Res.} A308, p.197, 1991,

\bibitem{14}
Surdo, A.I., Kortov, V.S. and Milman, I.I.,
{\em Ukr. Fiz. J.} 33, p.872,
1988 (Russia)

\bibitem{15}
Jeffries, B., Summers, G.P., Craword, J.H.,
$F$-center fluorescence in neutron-bombarded sapphire,
{\em J. Appl.
Phys.} 51, p.3984, 1980

\bibitem{16}
Harutunyan, V.V., Babayan, A.K., Gevorkyan, V.A.,
{\em FTT} v.37, p.443, 1995 (Russia)


\end{thebibliography}
\end{document}